\documentclass[a4paper,fleqn,usenatbib]{mnras}

\usepackage{mathptmx}

\usepackage[T1]{fontenc}
\usepackage{ae,aecompl}

\usepackage{graphicx}	
\usepackage{amsmath}	
\usepackage{amssymb}	
\usepackage{comment}

\usepackage{multirow, array}

\excludecomment{comment}
\newcommand{\filu}{\textit{u'}}	
\newcommand{\filg}{\textit{g'}}	
\newcommand{\filr}{\textit{r'}}	
\newcommand{\filus}{\textit{u$_s$}}	
\newcommand{\filgs}{\textit{g$_s$}}	
\newcommand{\filrs}{\textit{r$_s$}}	
\newcommand{\ufreq}{day$^{-1}$}	

\newcommand{\warwick}{$^{1}$}
\newcommand{\unc}{$^{2}$}
\newcommand{\mpi}{$^{3}$}
\newcommand{\tlu}{$^{5}$}
\newcommand{\sheffield}{$^{4}$}
\newcommand{\iac}{$^{7}$}
\newcommand{\ioacambs}{$^{6}$}

\newcommand{\ktwo}{\textit{K2}}
\newcommand{\Kepler}{\textit{Kepler}}

\newcommand{\longp}{2.1660(2)}
\newcommand{\ktop}{8.506(2)}
\newcommand{\ktsp}{6.339(2)}
\newcommand{\ktoph}{91.541(2)}
\newcommand{\ktsph}{89.374(2)}

\title[A 15.7-Minute AM\,CVn in \ktwo]{A 15.7-Minute AM\,CVn Binary Discovered in \textit{K2}}

\author[M. J. Green et al.]{M. J. Green\warwick\thanks{E-mail: matthew.green@warwick.ac.uk (MJG)},
J. J. Hermes\unc\thanks{Hubble Fellow},
T. R. Marsh\warwick,
D. T. H. Steeghs\warwick,
Keaton J. Bell\mpi,
\newauthor
S. P. Littlefair\sheffield, 
S. G. Parsons\sheffield,
E. Dennihy\unc,
J. T. Fuchs\tlu,
J. S. Reding\unc,
\newauthor
B. C. Kaiser\unc,
R. P. Ashley\warwick,
E. Breedt\ioacambs,
V. S. Dhillon\sheffield$^,$\iac,
N. P. Gentile Fusillo\warwick,
\newauthor
P. Kerry\sheffield,
and D. I. Sahman\sheffield.
\\
\warwick Astronomy and Astrophysics Group, Department of Physics, University of Warwick, Gibbet Hill Road, Coventry, CV4 7AL, UK
\\
\unc Department of Physics and Astronomy, University of North Carolina, Chapel Hill, NC 27599-3255, USA
\\
\mpi Max-Planck-Institut f{\"u}r Sonnensystemforschung, Justus-von-Liebig-Weg 3, 37077 G{\"o}ttingen, Germany
\\
\sheffield Department of Physics and Astronomy, University of Sheffield, Sheffield, S3~7RH, United Kingdom
\\
\tlu Department of Physics, Texas Lutheran University, Seguin, TX 78155
\\
\ioacambs Institute of Astronomy, University of Cambridge, Madingley Road, Cambridge, CB3~0HA, United Kingdom
\\
\iac Instituto de Astrof\'isica de Canarias, 38205 La Laguna, Tenerife, Spain
}

\date{Accepted XXX. Received YYY; in original form ZZZ}

\pubyear{2017}

\begin{document}
\label{firstpage}
\pagerange{\pageref{firstpage}--\pageref{lastpage}}
\maketitle

\begin{abstract}
We present the discovery of SDSS\,J135154.46-064309.0, a short-period variable observed using 30-minute cadence photometry in \ktwo\ Campaign 6. Follow-up spectroscopy and high-speed photometry support a classification as a new member of the rare class of ultracompact accreting binaries known as AM\,CVn stars. The spectroscopic orbital period of $15.65 \pm 0.12$\,minutes makes this system the fourth-shortest period AM\,CVn known, and the second system of this type to be discovered by the \Kepler\ spacecraft. The \ktwo\ data show photometric periods at $15.7306 \pm 0.0003$\,minutes, $16.1121 \pm 0.0004$\,minutes and $664.82 \pm 0.06$\,minutes, which we identify as the orbital period, superhump period, and disc precession period, respectively. From the superhump and orbital periods we estimate the binary mass ratio $q = M_2/M_1 = 0.111 \pm 0.005$, though this method of mass ratio determination may not be well calibrated for helium-dominated binaries.  This system is likely to be a bright foreground source of gravitational waves in the frequency range detectable by \textit{LISA}, and may be of use as a calibration source if future studies are able to constrain the masses of its stellar components.
\end{abstract}

\begin{keywords}
stars: individual: SDSS J135154.46-064309.0 -- stars: dwarf novae -- novae, cataclysmic variables -- binaries: close -- white dwarfs
\end{keywords}



\section{Introduction}
\label{introduction}

AM\,CVn-type systems are among the shortest-period binaries known, with orbital periods of 5--65~minutes. 
They are ultracompact binaries, consisting of a white dwarf accreting helium-dominated matter from a degenerate or semi-degenerate donor \citep[see][for recent reviews]{SolheimAMCVn,Breedt2015}. 
Their short orbital periods imply a small physical separation between the two stars.
Due to these small separations, AM\,CVns are among the brightest sources of gravitational waves in the frequency range that will be visible to the Laser Interferometer Space Antenna (\textit{LISA}). 
The shortest-period AM\,CVns have been suggested as calibration sources for \textit{LISA} \citep{Korol2017,Nelemans2004}. 
AM\,CVns are probes of helium accretion physics \citep{Kotko2012,Cannizzo2015a} and can be used to constrain the poorly-understood common envelope phase of compact binary evolution \citep{Ivanova2013}.

The majority of AM\,CVn systems are thought to begin mass transfer at orbital periods $\lesssim 15$\,minutes and evolve to longer periods throughout their lives \citep{Paczynski1967, Savonije1986, Iben1987, Deloye2007,Yungelson2008}. 
This evolution is driven primarily by the loss of angular momentum through gravitational wave radiation, which is strongest at short periods and declines steeply as the period increases. By tracking the period evolution of these binaries over timescales of years it is possible to use their gravitational wave radiation as a means to constrain the elusive masses of the component stars  \citep[eg.][]{deMiguel2018,Copperwheat2011a}. 

AM\,CVns span a wide range of accretion rates, from $10^{-7.5}~M_\odot~\text{yr}^{-1}$ at the shortest periods to $10^{-12}~M_\odot~\text{yr}^{-1}$ at long periods \citep{Deloye2007}. 
The behaviour of the accretion disc consequently changes. At short periods ($\lesssim 20$~minutes), the high accretion rate drives the accretion disc into a constant `high state' in which the disc is optically thick and dominates the optical flux from the system, comparable to nova-like cataclysmic variables (CVs).
Long period, low accretion rate AM\,CVn stars are conversely in a constant `low state' in which the disc is relatively faint and the white dwarf dominates the optical flux. Intermediate period AM\,CVn stars (20--50~minutes) alternate between low-state `quiescent' periods and high-state `outbursts', analogous to dwarf nova outbursts.

A fourth category of AM\,CVn stars exists, which contains the two shortest-period binaries known (HM\,Cnc and V407\,Vul, both with orbital periods less than 10 minutes). These systems do not seem to behave according to the high state model. Several alternate models for these systems have been proposed, of which the simplest is that they are in a state in which the accreted material impacts directly onto the surface of their central white dwarfs, as the compactness of these systems prohibits the formation of accretion discs \citep{Marsh2002a,Roelofs2010}. A third binary, ES\,Cet, may also belong to this category \citep{Espaillat2005}, but this has not been confirmed. In this work we will treat ES\,Cet as a high-state disc system.

\begin{figure}
\includegraphics[width=\columnwidth]{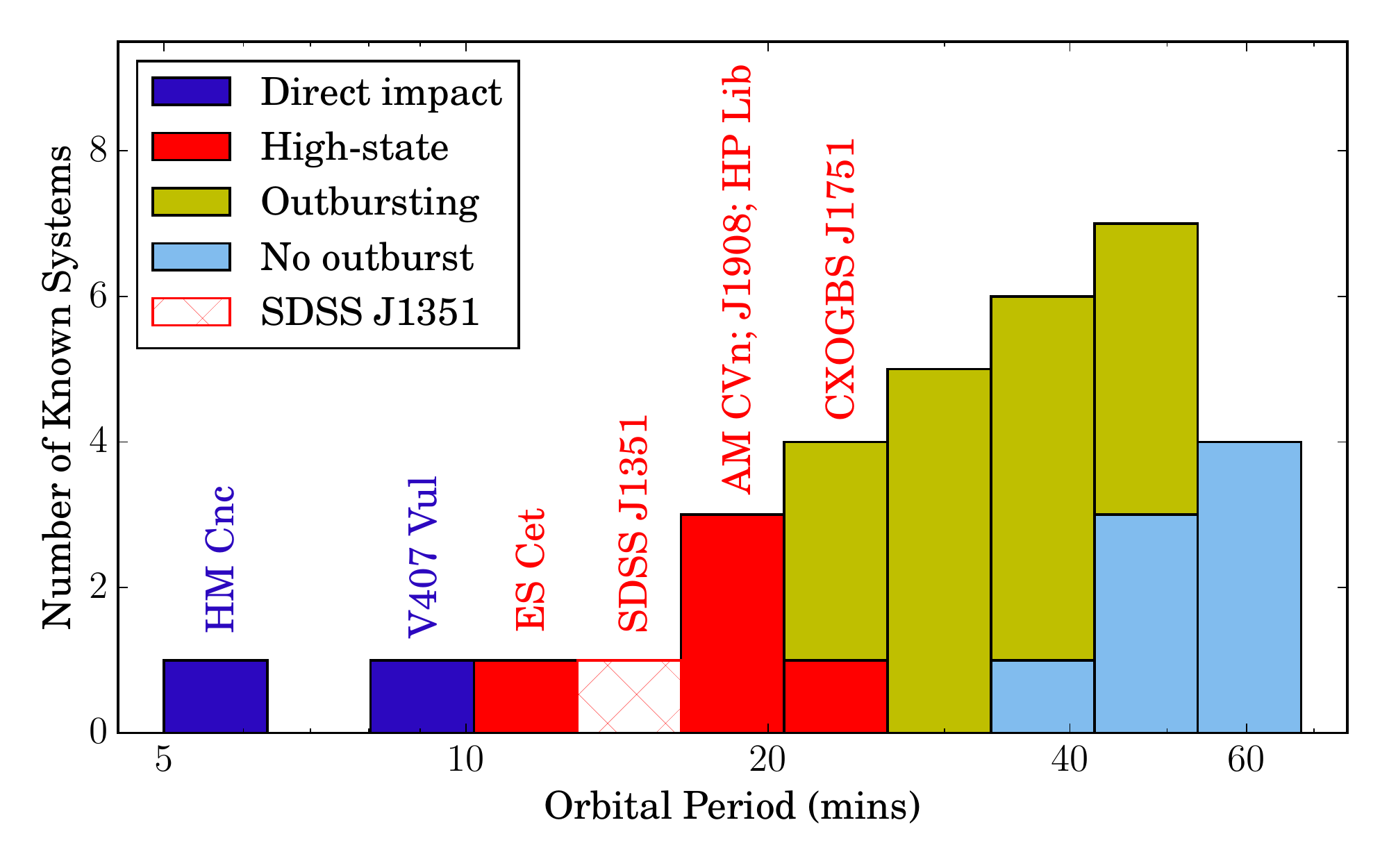}
\caption{The logarithmic orbital period distribution of all AM\,CVns with published orbital periods, highlighting the orbital period of J1351 presented in this work. High-state and direct-impact binaries are labelled. The periods of these systems, along with references, are given in Table~\ref{tab:amcvns-j1351}.
}
\label{fig:histogram}
\end{figure}

The period distribution of AM\,CVn stars is shown in Figure~\ref{fig:histogram}, and their periods are summarised in Table~\ref{tab:amcvns-j1351}.
Owing to the high rate of period change at short periods, high-state AM\,CVn stars are expected to be in the minority \citep{Deloye2007}.
Only 5 disc-accreting, high-state AM\,CVn-type systems are currently known (including ES\,Cet). 
There is a large gap at short periods between ES\,Cet (10.3~minutes) and AM\,CVn itself (17.1~minutes).

Although AM\,CVn stars often show variability on a multitude of timescales \citep{Fontaine2011,Kupfer2015}, there are three characteristic timescales that have physical motivation \citep{Skillman1999}. Firstly, the orbital period can be measured spectroscopically, and in some systems has a photometric equivalent as well \citep[eg.][]{Copperwheat2011a}. Secondly, if the disc of the AM\,CVn is eccentric \citep[as is possible due to their mass ratios,][]{Whitehurst1988}, the disc will precess under the tidal field of the donor. This precession period is occasionally visible in either spectroscopy or photometry of AM~CVn systems \citep[eg.][]{Patterson1993,Skillman1999}. Thirdly, a photometric signal at a period known as the `superhump' period is visible in many AM\,CVn stars, especially in high-state systems or systems in outburst. This signal originates from a tidal interaction between the disc and the donor star, and is found at the beat frequency between the orbital and disc precession periods \citep{Patterson1993}
\begin{equation}
f_{\text{sh}} = f_{\text{orb}} \pm f_{\text{prec}}.
\label{eq:freqs-j1351}
\end{equation}
A superhump period which is longer than the orbital period (`$-$' in Equation~\ref{eq:freqs-j1351}) indicates that the disc precession is apsidal (precession within the plane of the system). A superhump period which is shorter than the orbital period indicates that the disc precession is nodal (precession of the axis of rotation of a tilted disc). In both AM\,CVn binaries and SU\,UMa binaries (a class of CVs which exhibit the same phenonemon), apsidal precession is found to be more common. The orbital period and superhump period are similar in length (generally within a few percent). Therefore, even in systems which show photometric signatures on both timescales, photometry over a long baseline is often required to separate the two signals \citep[eg.][]{Armstrong2012}. 

Space-based photometry can be a powerful tool for resolving similar signals by providing continuous, long-baseline coverage of a target. \citet{Fontaine2011} reported the discovery of SDSS~J1908+3940, a high-state AM\,CVn found in the \Kepler\ field. 
The full 1052-day \Kepler\ lightcurve on that system was presented in \citet{Kupfer2015}, in which the long baseline allowed for exquisite constraints on the system's periods and their long-term phase evolution.
In this paper, we present the discovery of SDSS\,J135154.46-064309.0 (henceforth J1351), a high-state AM\,CVn discovered in long-cadence \ktwo\ photometry from Campaign~6. 

In Section~\ref{observations}, we describe the original \ktwo\ observations as well as follow-up observations undertaken to characterise the system. In Section~\ref{data}, we present the data obtained during these observations. Finally in Section~\ref{discussion}, we justify the AM\,CVn classification and describe our interpretation of these data in the context of that classification.

\section{Observations}
\label{observations}

A summary of the observations obtained for this work is given in Table~\ref{tab:observations}.

\begin{table}
	\centering
	\caption{A summary of the follow-up observations presented in this work. Spectra are in the wavelength range 3600--5200~\AA\ except where marked with a $\dagger$.
	}
	\label{tab:observations}
	\begin{tabular}{lccr} 
		\hline
		Instrument & Date & \multirow{2}{6.5em}{\centering Filters / \\Slit Width (")} & \multirow{2}{4em}{\raggedleft Exposure\\ time (s)}\\
		&&&\\
		\hline
		McDonald & 2017-03-02 & BG40 & 241$\times$30\\
		McDonald & 2017-03-03 & BG40 & 101$\times$30\\
		McDonald & 2017-03-05 & BG40 & 324$\times$30\\
		McDonald & 2017-03-06 & BG40 & 595$\times$30\\
		SOAR spectra & 2017-04-20 & 3.21 & 6$\times$300\\
		SOAR spectra & 2017-04-20 & 1.19 & 8$\times$300\\
		SOAR phot. & 2017-04-20 & S8612 & 125$\times$20\\
		SOAR spectra & 2017-04-21 & 3.21 & 4$\times$300\\
 		SOAR phot. & 2017-04-21 & S8612 & 230$\times$20\\       
		ULTRACAM & 2017-05-03 & \filu\filg\filr & 780$\times$8 *\\
		ULTRACAM & 2017-05-04 & \filu\filg\filr & 865$\times$8 *\\
		ULTRACAM & 2017-05-14 & \filus\filgs\filrs & 1516$\times$7 *\\
		SOAR spectra$\dagger$ & 2017-05-29 & 1.19 & 4$\times$300\\
		SOAR spectra & 2017-05-29 & 1.19 & 6$\times$300\\
		SOAR spectra & 2017-05-30 & 1.19 & 24$\times$300\\		
		\hline
	\multicolumn{4}{p{8cm}}{* Exposure times for \filu\ and \filus\ were increased by a factor of 3 to compensate for the lower sensitivity in that band.}\\
		\hline
	\multicolumn{4}{p{8cm}}{$\dagger$ Wavelength range 5200--6700\AA.}\\
		\hline
	\end{tabular}
\end{table}

\subsection{\ktwo\ Photometry}

J1351 (a.k.a. EPIC\,212759353, $K_p=18.9$\,mag) was observed in \ktwo\ Campaign 6, which lasted roughly 80 days from 14 July 2015 to 30 September 2015.
J1351 was targeted with long-cadence (29.4-min) exposures as a high-probability white dwarf candidate, based on its blue colours and high proper motion \citep{GentileFusillo15}.

We examined several different pipeline extractions of the \ktwo\ photometry, and settled on a final light curve produced from the Pre-search Data Conditioning pipeline from the \Kepler\ Guest Observer office \citep{VanCleve16}, which uses a 4-pixel fixed aperture. 

\subsection{McDonald/ProEM Photometry}

The appearance of periodic variability in the \ktwo\ data of J1351 motivated the collection of follow-up data to characterise the system. We obtained time series photometry on J1351 on 2017 March 2, 3, 5, and 6 with a frame-transfer Princeton Instruments ProEM camera on the 2.1m Otto Struve Telescope at McDonald Observatory. Observations were made through a broad ($3300-6000$\,\AA) BG40 filter to reduce sky noise, and each exposure was 30\,s long. We dark- and flat-corrected each frame with standard IRAF tasks, using calibration data from the start of each observing night. We measured circular aperture photometry for the target and two bright comparison stars in the field using the IRAF script \texttt{CCD\_HSP} \citep{Kanaan2002}.  
We correct for transparency variations and obtain our final relative light curves by dividing the target flux by the weighted mean of the comparison star fluxes.

\subsection{SOAR/Goodman Spectroscopy}

We obtained optical spectra in April and May 2017 using the 4.1-m Southern Astrophysical Research (SOAR) telescope at Cerro Pach\'{o}n in Chile. We used the high-throughput Goodman spectrograph \citep{clemens04} with a 930 line mm$^{-1}$ grating, with two different grating/camera angle setups that cover roughly $3600-5200${\AA} and $5200-6700${\AA}.
In April 2017 we used both a 3.21\arcsec\ and 1.19\arcsec\ slit, covering a wavelength range of $3600-5200${\AA}. In May 2017, the 1.19\arcsec\ slit was used in both wavelength ranges.
The 1.19\arcsec\ spectra have a resolution of 2.4\,\AA.

All spectra were reduced using the software packages \textsc{pamela} and \textsc{molly}. 
The data obtained with the 1.19\arcsec\ slit were wavelength calibrated using iron arc lamp spectra that were recorded before and after the spectra, as well as being interspersed every 30~minutes on 30 May. The 3.21\arcsec\ spectra were wavelength-calibrated using a master arc that was recorded prior to observing, but these spectra suffer from large wavelength drifts and are not reliable for precision velocities.

The April spectra were flux-calibrated using the standard star LTT 3218, observed through a 3.21\arcsec\ slit. No flux standard was observed in May due to poor weather conditions. Instead, these spectra were flux-calibrated by comparison with the April data. A third-order spline was fitted to averaged spectra from each night, and uncalibrated spectra were multiplied throughout by the ratio of those splines.

\begin{figure*}
\includegraphics[width=500pt]{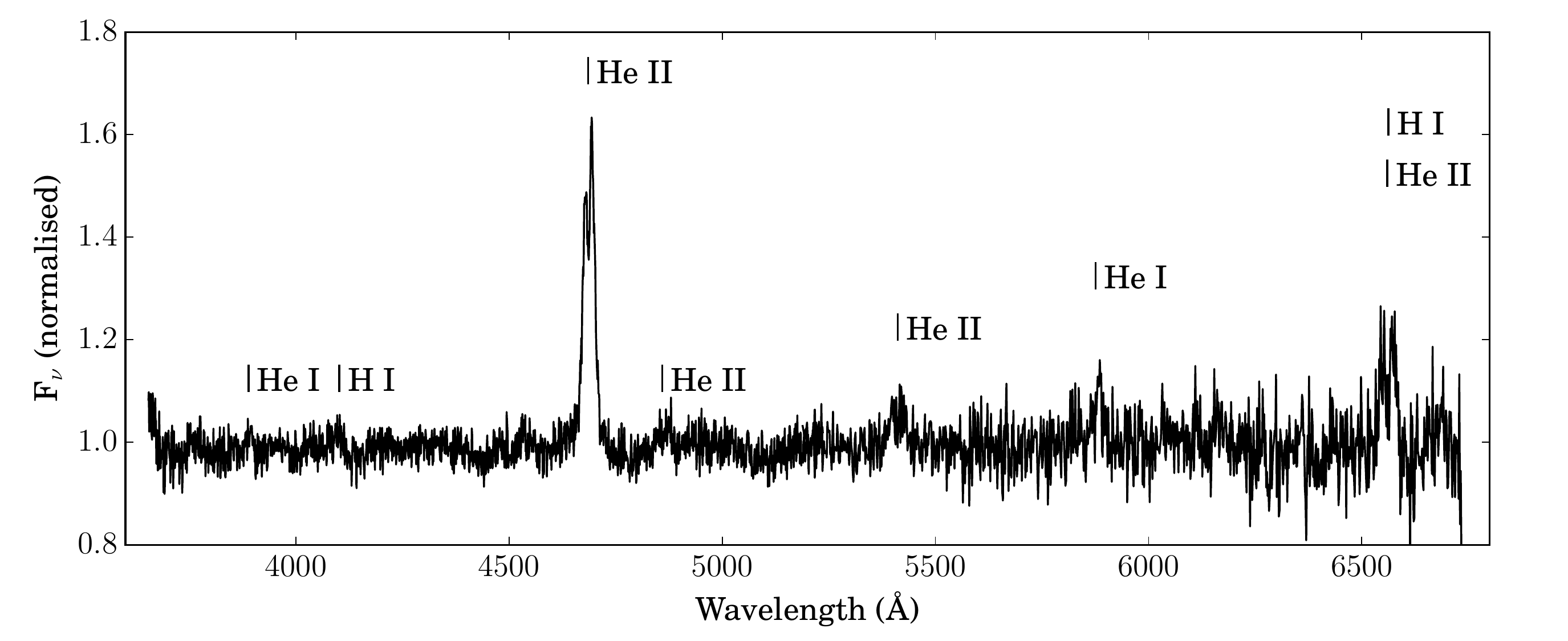}
\caption{SOAR spectrum of J1351, showing strong, double-peaked HeII emission at 4686~\AA\ and almost no emission or absorption of other elements. This spectrum is the average of 38 spectra in the 3000-5200~\AA\ range (all of those with a slit width of 1.3"), and 4 spectra in the 5200-6700~\AA\ range. The spectrum in the 5200-6700~\AA\ range has been rebinned by a factor of 2 for visualisation purposes. The spectrum has been normalised by dividing by the continuum, which increases the apparent strength of lines at longer wavelengths where the continuum is weaker. 
}
\label{fig:spectrum-j1351}
\end{figure*}

\begin{figure}
\includegraphics[width=\columnwidth]{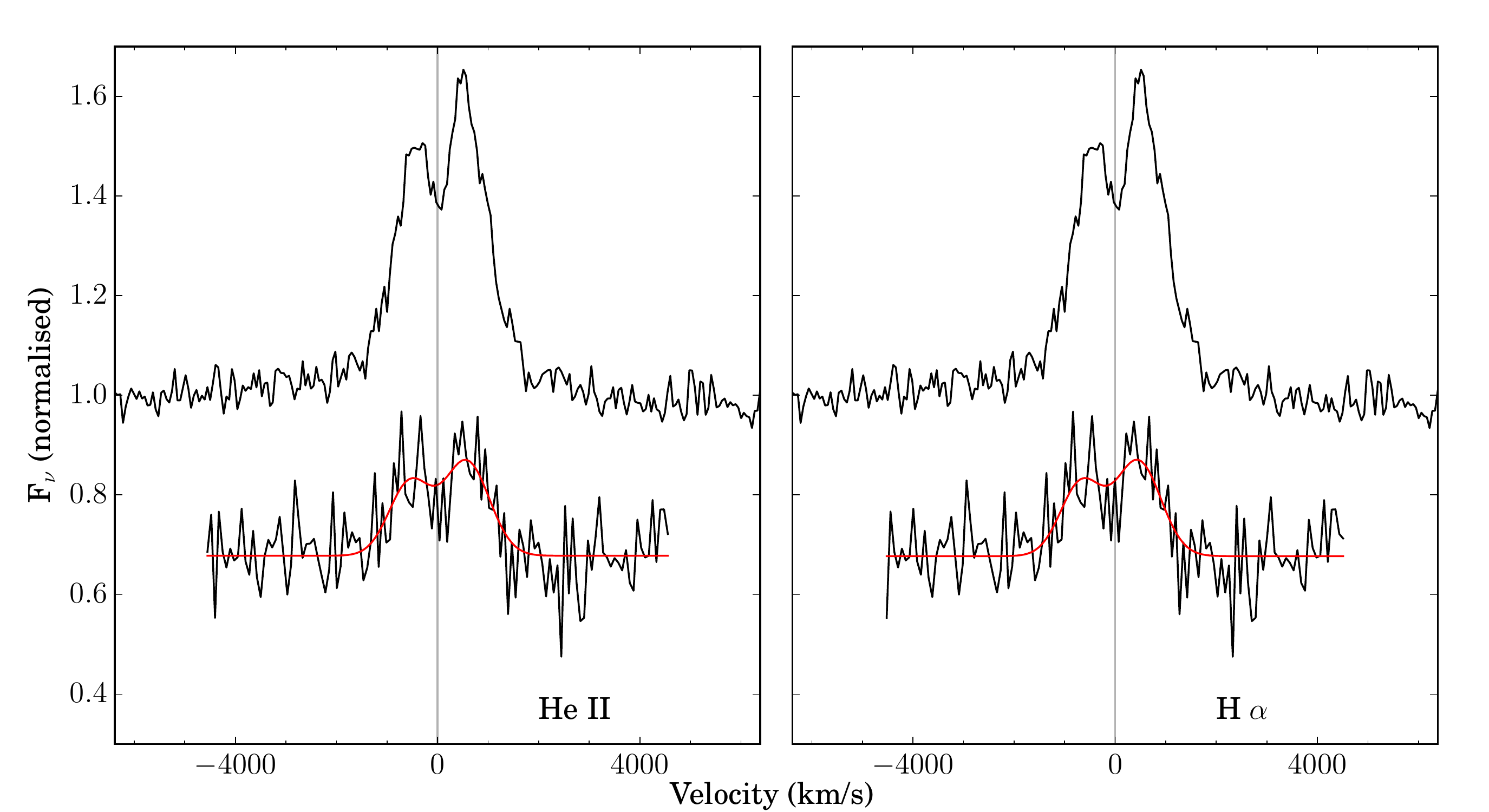}
\caption{Emission lines of J1351 at 4686\,\AA\ (above) and 6560\,\AA\ (below, offset vertically by -0.3), in both cases converted to velocity space. In the left panel, the 6560\,\AA\ line was converted assuming it is a He~II line with a central wavelength of 6560.10\,\AA. In the right panel, it is assumed to be the H$\alpha$ line with a central wavelength of 6562.72\,\AA. 
The red line shows a double Gaussian fit to the 6560\,\AA\ line, described in Section~\ref{spec}.
In both cases this line appears to be slightly blue-shifted relative to the 4686\,\AA\ line, but the discrepancy is more significant when the line is treated as H$\alpha$.
}
\label{fig:halpha}
\end{figure}

\begin{figure}
\includegraphics[width=\columnwidth]{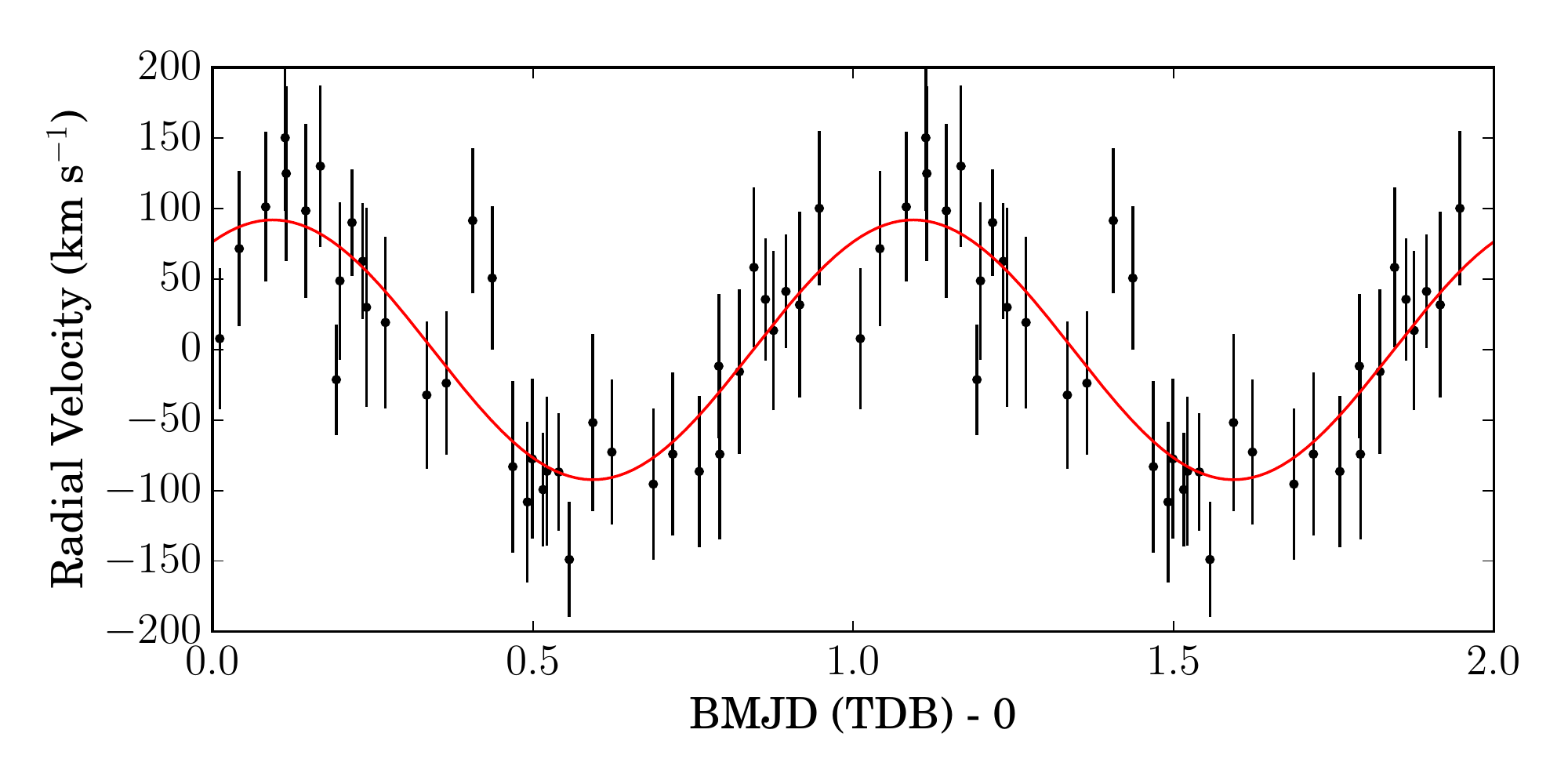}
\caption{RV data for J1351, phase-folded on a frequency of 91.7~\ufreq. This figure includes RV measurements from all 38 spectra that were observed with a 1.31" slit and which include the 4686~\AA\ line.
}
\label{fig:rvs-j1351}
\end{figure}

\subsection{SOAR/Goodman Photometry}

We also followed up J1351 with time-series photometry using SOAR/Goodman over two consecutive nights in April 2017. Our observations were obtained through a blue, broad-bandpass, red-cutoff S8612 filter. All exposures were 20\,s, with roughly 2.1\,s dead time for readouts. We bias- and flat-corrected each frame with standard IRAF tasks, and performed circular aperture photometry. 

\subsection{NTT/ULTRACAM Photometry}

Further follow-up photometry was obtained using ULTRACAM, a high-speed, triple-band photometer which uses frame-transfer CCDs to reduce the readout time overhead to negligible amounts \citep[25~ms; see][for a full description of the instrument]{ultracam}. For these observations ULTRACAM was mounted on the 3.5\,m New Technology Telescope (NTT) at the La Silla Observatory in Chile. 
J1351 was observed in May 2017 using Sloan \filu\filg\filr\ filters for two nights and the custom `Super-SDSS' filters \filus\filgs\filrs\ for the third night. The latter set of filters are designed to cover the same wavelength range as \filu\filg\filr\ filters with a higher throughput. 

The ULTRACAM data were reduced using the standard ULTRACAM pipeline. Images were bias- and dark-subtracted and were divided throughout by a flat field taken during the run. Due to poor weather conditions no flat field was available using the \filus\filgs\filrs\ filters, so \filu\filg\filr\ flats were used instead. The target was flux-calibrated using a nearby, non-variable SDSS comparison star (SDSS J135203.48-064405.1, $m_\filu = 17.26$, $m_\filg = 15.64$, $m_\filr = 15.11$, all error bars 0.01~mag or less). As no flux standards exist for the \filus\filgs\filrs\ filters the absolute flux calibration of data from those filters may not be reliable, but these data are only used for timing purposes. Transparency changes due to clouds or atmospheric thickness were removed using the same comparison star.

\section{Analysis}
\label{data}

\begin{figure}
\includegraphics[width=\columnwidth]{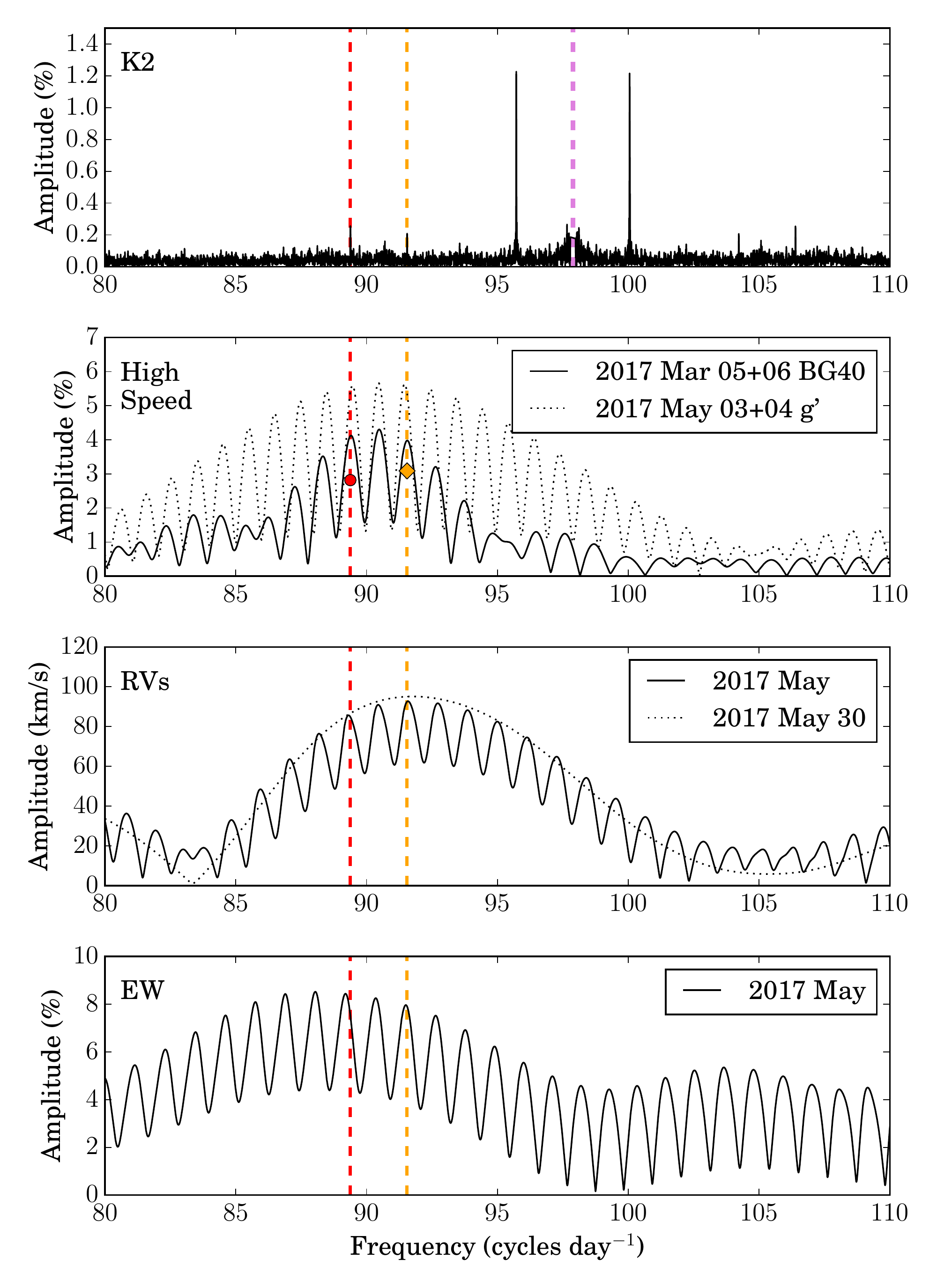}
\caption{Lomb-Scargle periodograms of various datasets overplotted for comparison. 
Red and yellow dashed lines show the proposed superhump and orbital periods, with circle and diamond markers in the second panel indicating the predicted amplitudes of those signals (calculated from the \ktwo\ signals according to Equation~\ref{eq:smearing}, and assuming the amplitude of variation is constant for all wavelengths of light). The purple dashed line shows the harmonic of the \ktwo\ Nyquist frequency.
The strong \ktwo\ peaks in the top panel are Nyquist bounces of the \longp\ peak, which clearly have no corresponding peak in the high-cadence data.
}
\label{fig:pgram}
\end{figure}

\begin{figure}
\includegraphics[width=\columnwidth]{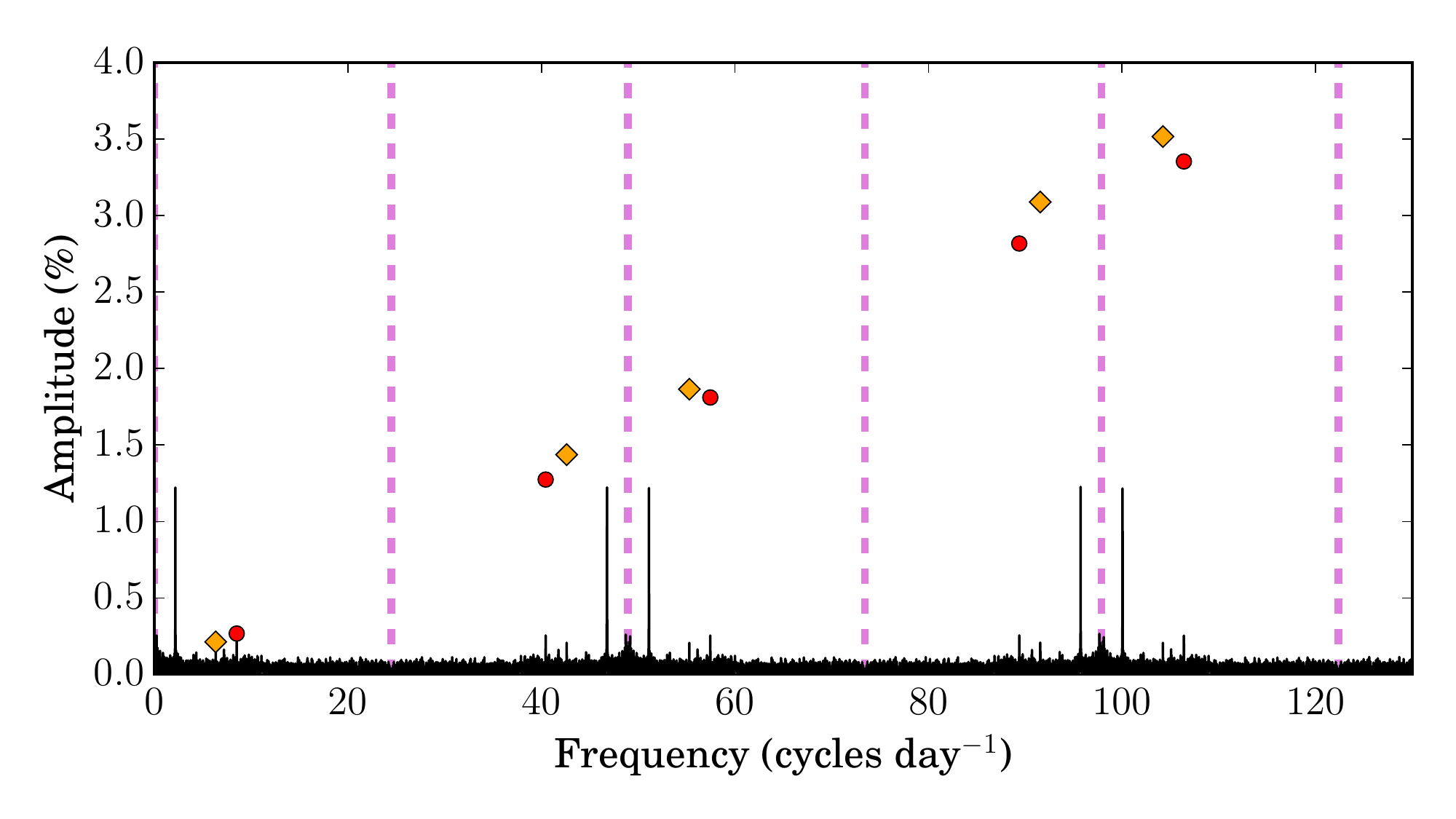}
\caption{Lomb-Scargle periodogram of K2 data, showing a clear peak at 2.162(4)\,\ufreq\ along with its higher-frequency Nyquist bounces. The Nyquist frequency of 24.47\,\ufreq\ and its harmonics are shown by dashed purple lines. 
Red circles and yellow diamonds show the predicted frequencies and intrinsic amplitudes of super-Nyquist signals that may correspond to the \ktop\,\ufreq\ and \ktsp\,\ufreq\ \ktwo\ signals, with frequencies reflected about the Nyquist frequency and its harmonics, and amplitudes corrected according to Equation~\ref{eq:smearing}.
}
\label{fig:k2-ls}
\end{figure}

\begin{figure}
\includegraphics[width=\columnwidth]{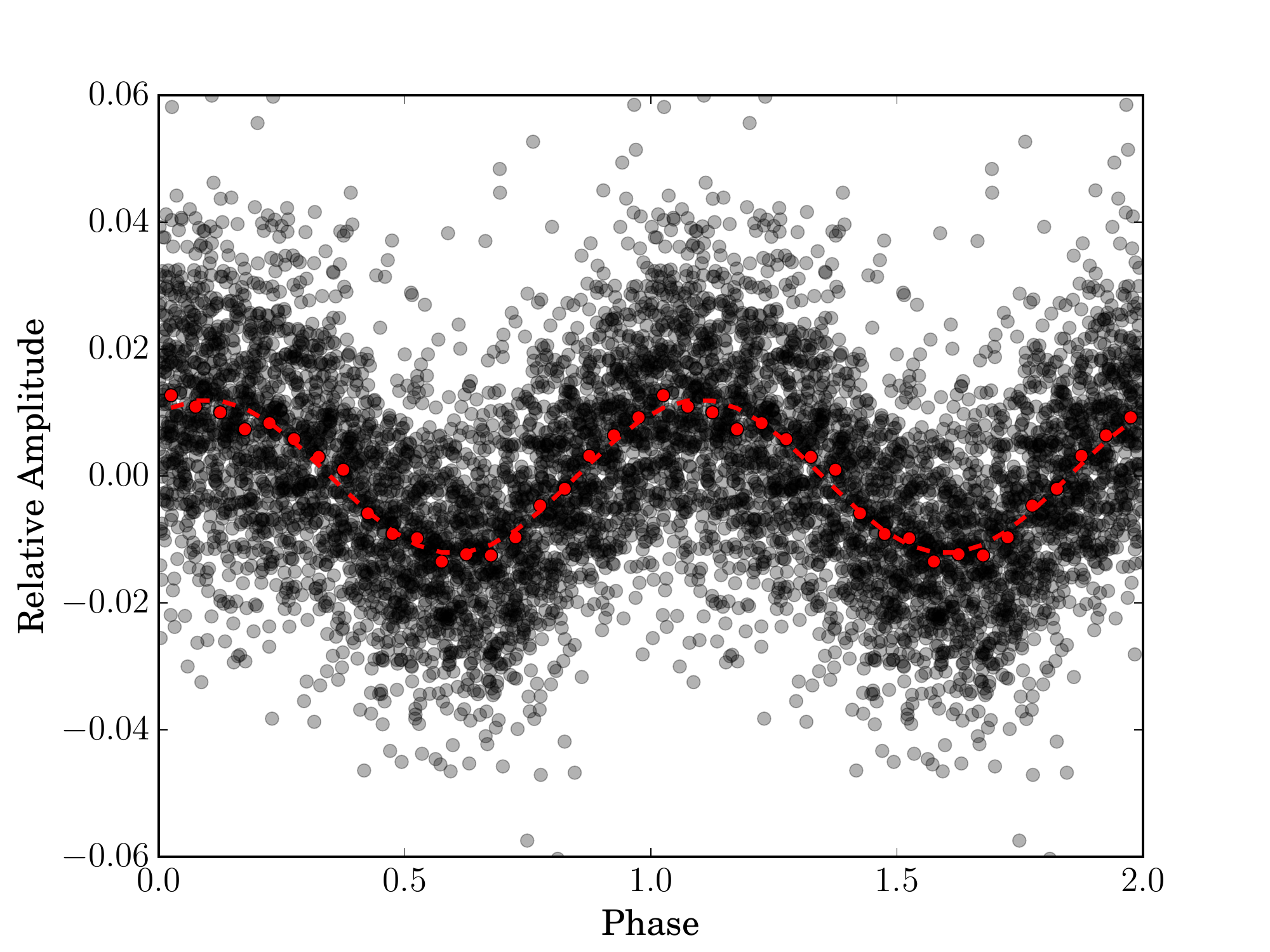}
\caption{Phase-fold of the \ktwo\ data on a frequency of 2.162~\ufreq. Individual data are shown as black circles, with red circles representing 20 bins of the data. Although the individual data show considerable scatter, the binned data is well-described by a sinusoid (red dashed line). 
}
\label{fig:k2-phasefold-j1351}
\end{figure}

\begin{figure}
\includegraphics[width=\columnwidth]{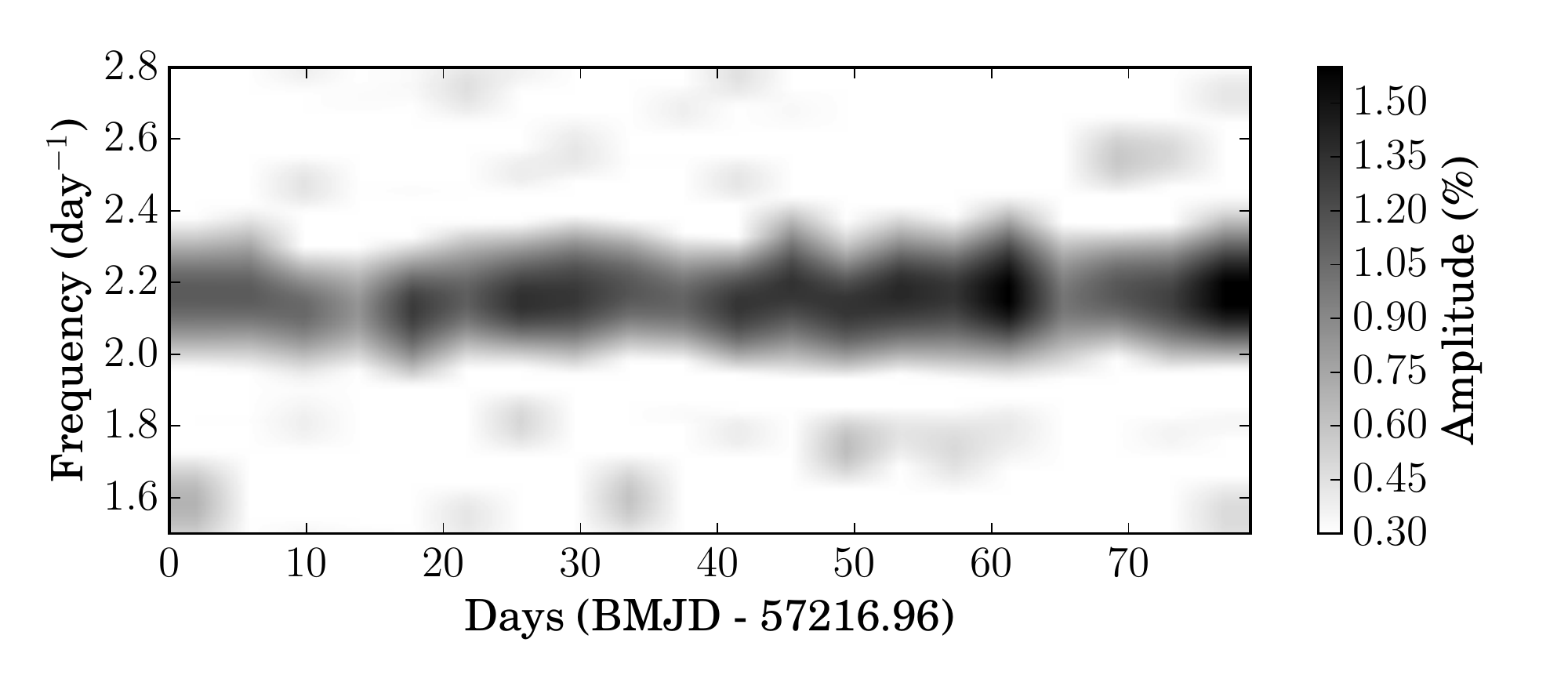}
\caption{A running periodogram of the \ktwo\ data, created by separating the \ktwo\ data into 20 non-overlapping subsections and calculating periodograms of each. We find no significant variation in the  frequency of the \longp\,\ufreq\ signal.
}
\label{fig:running-ls}
\end{figure}

\begin{figure*}
\includegraphics[width=500pt]{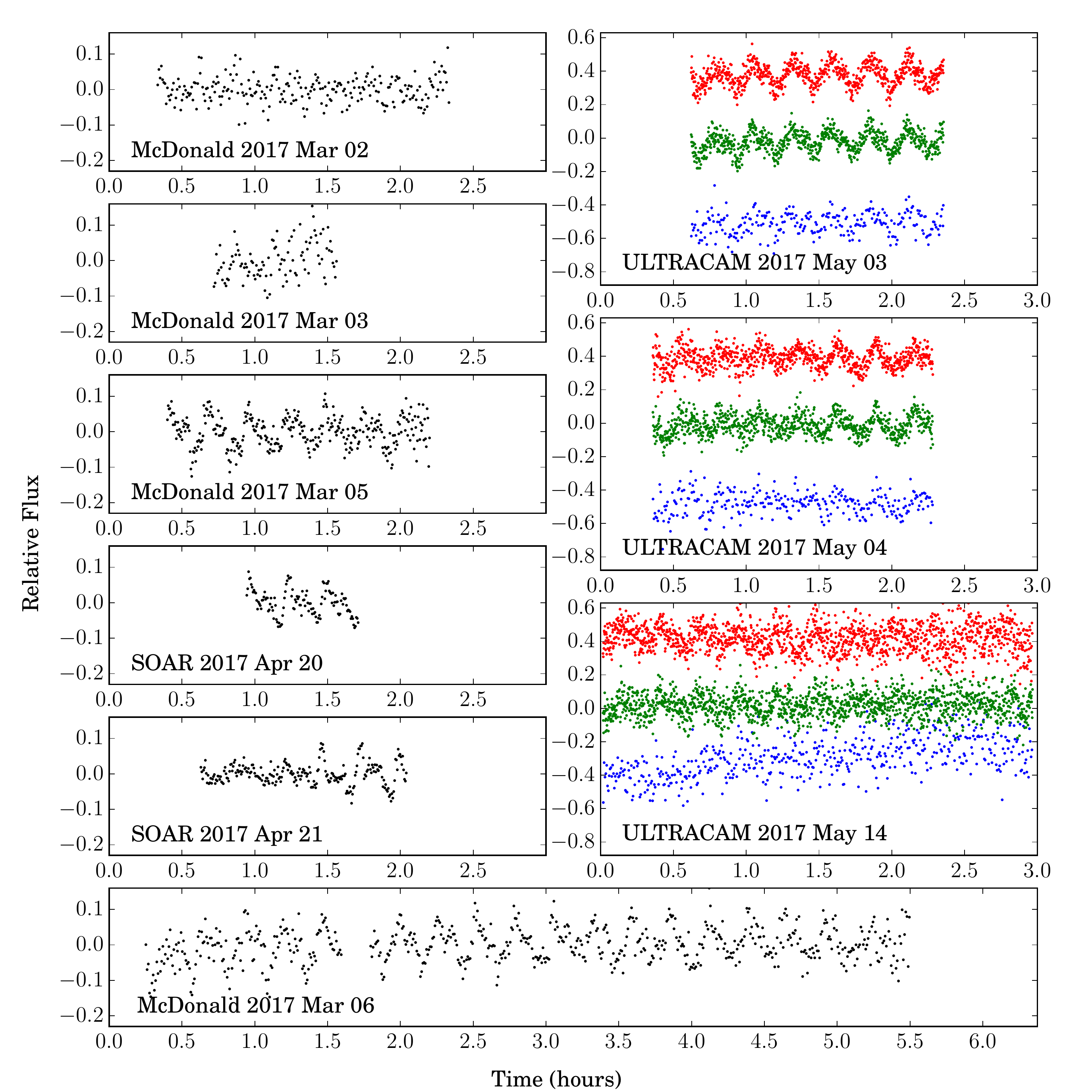}
\caption{All short-cadence photometry of J1351 obtained for this project. For the ULTRACAM data, the \filr/\filrs-band and \filu/\filus-band data have been offset by +0.4 and -0.4 respectively. Most observing runs clearly show sawtooth-shaped variability on a frequency of around 90~\ufreq. On some nights (see April 21) this signal changes significantly in shape and amplitude between cycles. Some runs (see March 06 and May 03) also show evidence for longer-term variability in brightness. The quality of data from March 02, March 03 and May 14 data was affected by poor conditions.
}
\label{fig:allphotometry}
\end{figure*}

\subsection{Spectroscopy}
\label{spec}

The SOAR spectra of J1351 (Figure~\ref{fig:spectrum-j1351}) show a clear double-peaked emission line at 4686\,\AA, consistent with He~II emission. He~II emission lines are also present at 4859\,\AA, 5412\,\AA\ and 6560~\AA, corresponding to the Pickering series, though the latter of these occurs at a similar wavelength to H$\alpha$. Weak He~I emission lines are seen at 3889\,\AA\ and 5876\,\AA. A weak feature is also seen at 4102~\AA\ which may correspond to hydrogen, but is difficult to confirm at this S/N. 

The lack of significant H$\beta$ or H$\gamma$ suggests that the feature at 6560\,\AA\ can be attributed to He~II rather than H$\alpha$. 
The strength of the 6560\,\AA\ line is comparable to the strengths of the other lines in the Pickering series. 
In order to provide an independent test of this identification, we converted that region of the spectrum to velocity space twice. In the first conversion we used a central wavelength corresponding to that of the He~II line, and in the second to that of H$\alpha$. These conversions can be compared to the velocity of the He~II 4686\,\AA\ to identify the closest match (Figure~\ref{fig:halpha}). By fitting the line profiles with a double Gaussian shape (first converged on the He~II 4686\,\AA\ line to constrain its parameters), we measured velocity shifts of $-20 \pm 70$\,km/s for the He~II case and $-140 \pm 70$\,km/s for the H$\alpha$ case. This gives a marginal preference for He~II, in accordance with our identification.

We measured the radial velocity (RV) shift of the 4686\,\AA\ line in each spectrum using a double-Gaussian fit. The RVs vary on a short-period sinusoidal pattern, as shown in a phase-fold of the measured RVs (Figure~\ref{fig:rvs-j1351}). We measured this frequency by fitting a sine wave to the data. We found that the best results came from fitting to the 30 May data alone, this being the longest stretch of continuous data. The resulting frequency is $92.0 \pm 0.7$\,\ufreq. 
A Lomb-Scargle periodogram \citep[as implemented in the Python package \texttt{astropy}]{Lomb,Scargle,VanderPlas2017} combining consecutive nights of RV data splits this peak into multiple aliases (Figure~\ref{fig:pgram}).

The 4686\,\AA\ line has an equivalent width of $-16.6 \pm 0.2$\,\AA. There is some periodicity in the equivalent widths measured (Figure~\ref{fig:pgram}). However, its period is closer to the observed photometric period (next section) than the spectroscopic RV period, and we suggest that this results from variations in the continuum rather than in the spectral line.

\subsection{\ktwo\ Photometry}
\label{k2photom}

\begin{table}
	\centering
	\caption{The sub-Nyquist peaks in a periodogram of \ktwo\ data. Frequencies and amplitudes were found by fitting sine waves to the \ktwo\ data. 
	}
	\label{tab:k2frequencies}
	\begin{tabular}{lc}
		\hline
		Frequency (day$^{-1}$) & Measured Amplitude (\%)\\
		\hline
		\longp & 1.20(4) \\
        \ktop & 0.21(5) \\
        \ktsp & 0.26(7) \\

        
        
		\hline
	\end{tabular}
\end{table}

The Nyquist frequency of the \ktwo\ data is 24.47\,\ufreq. For any signal with a frequency higher than the Nyquist frequency of the data, there are several systematic effects which must be taken into account \citep{Bell2017}. 
Firstly, any super-Nyquist frequency will be under-sampled, and therefore there will be a sub-Nyquist frequency from which it is indistinguishable. This results in the effect known as `Nyquist bounces', in which a periodogram will show each signal several times, reflected off each harmonic of the Nyquist frequency. 
Secondly, as the exposure time of the observation is significant compared to the period of any super-Nyquist variability, the signal will be smeared. This causes a reduction in the measured amplitude. For a sinusoidal signal, this reduction in amplitude can be described by 
\begin{equation}
A_\text{measured} / A_\text{intrinsic} = \text{sinc} (\pi t_\text{exp} / P )
\label{eq:smearing}
\end{equation}
where $P$ is the period of the signal, $t_\text{exp}$ is the exposure time, and $\text{sinc}(x) = \sin(x)/x$.

The \ktwo\ data show a signal at a frequency of \longp\,\ufreq. A Lomb-Scargle periodogram of these data shows that this signal and its Nyquist bounces have by far the highest measured amplitudes of all signals present (Figure~\ref{fig:k2-ls}). 
A phase fold of the \ktwo\ photometry on the \longp\,\ufreq\ frequency shows that the signal is well-approximated by a sinusoid (Figure~\ref{fig:k2-phasefold-j1351}). 
Given the smearing effect described in Equation~\ref{eq:smearing}, the intrinsic amplitude of this variability would have to be large ($> 25 \%$) if the intrinsic frequency were super-Nyquist. 
To investigate the constancy of this signal, we separated the \ktwo\ data into 20 non-overlapping sections and produced a periodogram of each section (Figure~\ref{fig:running-ls}). When compared with a constant frequency the frequencies measured from these subdivisions have $\chi^2 = 24.1$, and the phase offsets have $\chi^2 = 16.2$, both for 19 degrees of freedom. We therefore do not find any evidence that the frequency of this signal is variable.

Two lower-amplitude signals are also visible in the periodogram (Figure~\ref{fig:k2-ls}). We summarise their sub-Nyquist frequencies in Table~\ref{tab:k2frequencies}. If these signals are super-Nyquist, their intrinsic amplitudes can be predicted by Equation~\ref{eq:smearing} (see Figure~\ref{fig:k2-ls}). 

These three signals obey Equation~\ref{eq:freqs-j1351} to within $1\sigma$. The same is true for any Nyquist bounce of the 8.506(2)\,\ufreq\ and 6.339(2)\,\ufreq\ signals, provided that both are subject to the same number of Nyquist bounces and that the \longp\,\ufreq\ signal is sub-Nyquist. In order to determine the number of Nyquist bounces which these signals are subject to, it is necessary to examine higher-cadence photometry of the system.


\subsection{High-cadence photometry}

\begin{figure}
\includegraphics[width=\columnwidth]{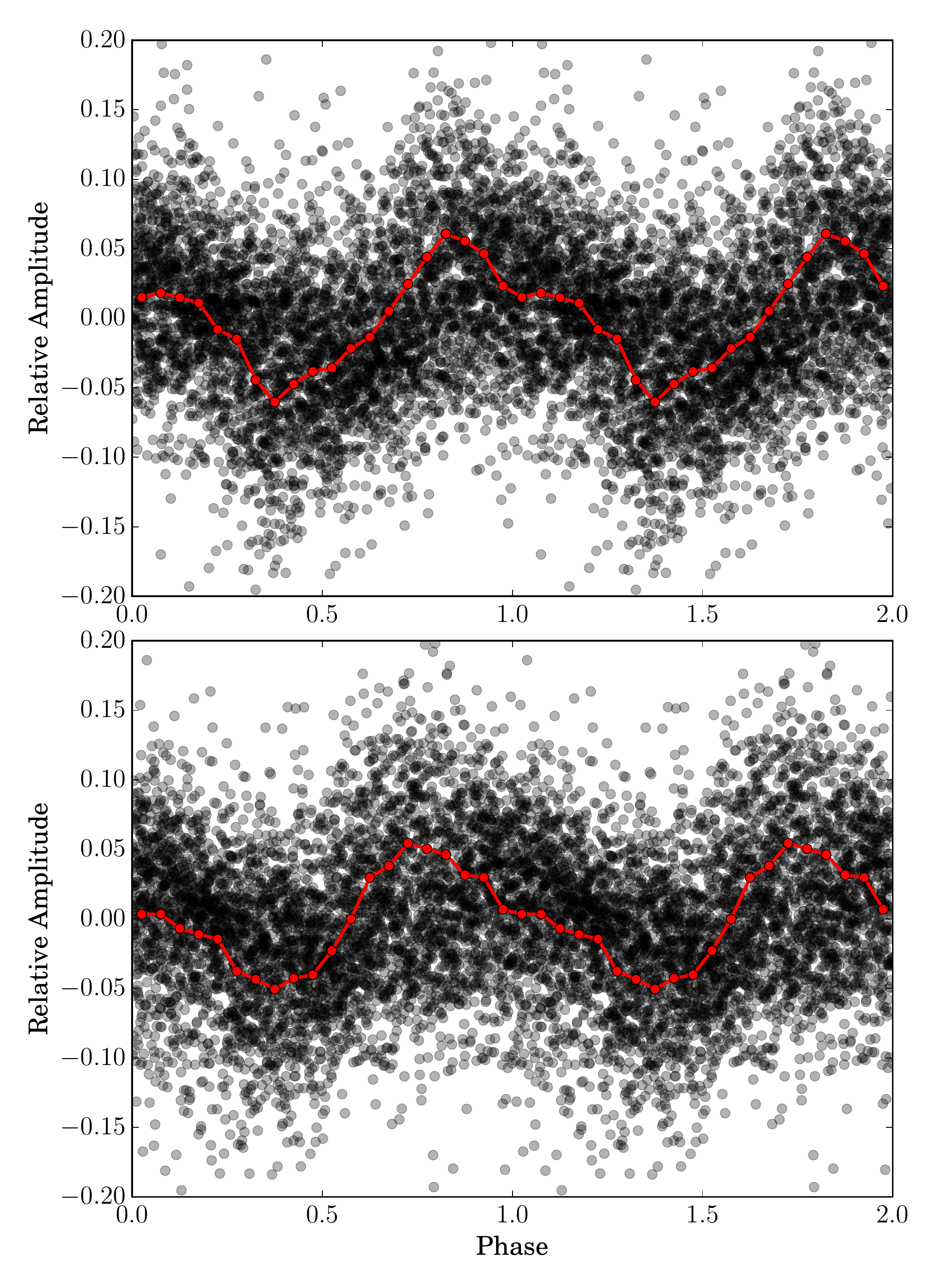}
\caption{Phase folds of all \filg\ ULTRACAM photometry and BG40 McDonald photometry of J1351 on frequencies of \ktoph\,\ufreq\ (top panel) and \ktsph\,\ufreq\ (bottom panel). The red lines show mean values calculated from a series of 20 phase bins. 
Both signals have a sawtooth shape, with a steep rise and a gentle decline, which may be the result of beating between the two signals. 
}
\label{fig:photom-phasefold-j1351}
\end{figure}


		
		
		
		
		
		
		
		

The McDonald, SOAR, and ULTRACAM photometric data are shown in Figure~\ref{fig:allphotometry}. The mean AB magnitudes and scatters across both ULTRACAM nights with \filu\filg\filr\ filters are $m_\filu = 18.46 \pm 0.08$, $m_\filg = 18.60 \pm 0.07$, and $m_\filr = 18.95 \pm 0.07$, where the quoted error bars are one standard deviation of the data so as to include the intrinsic variability of the system. The data show variability with a period of approximately 15\,minutes ($\approx$90\,\ufreq) and an amplitude on the order of 4--6\%, with changes to the lightcurve shape between one cycle and the next. Figure~\ref{fig:pgram} shows Lomb-Scargle periodograms produced from the ULTRACAM and McDonald data, each using data from two consecutive nights. Several nights also show long-term trends in brightness (on timescales longer than the observing window of these data), which may correspond to the \longp\,\ufreq\ signal, but are more likely to be due to changes in airmass as has previously been observed with the same McDonald setup \citep{Bell2017a}.

These data can be used to select between the Nyquist bounces of the signals in the \ktwo\ data discussed in Section~\ref{k2photom}. 
All Nyquist reflections of the \longp\,\ufreq\ peak in the \ktwo\ data can be easily ruled out. Given the smearing effect described in Equation~\ref{eq:smearing}, the intrinsic amplitude of this signal would have to be $> 25$\,\% if it were super-Nyquist, and such a signal is clearly not present in the short cadence photometry.
It is therefore most likely that the true frequency detected by K2 is \longp\,\ufreq, corresponding to a period of $664.82 \pm 0.06$\,min. This period is not measurable in the ground-based data due to the short observing windows of those data.

The $\approx$ 90\,\ufreq\ signal in the high-cadence data lies between the third Nyquist reflections of the measured frequencies for both the \ktsp\,\ufreq\ and the \ktop\,\ufreq\ signals, which would give these signals intrinsic frequencies of \ktsph\,\ufreq\ and \ktoph\,\ufreq. We therefore interpret the variability seen in the high-cadence data as a combination of both intrinsic signals. As shown in Figure~\ref{fig:pgram}, nightly aliasing makes it nearly impossible to disentangle the two signals using single-site, ground-based data. The amplitude measured in the high-cadence data agrees well with the prediction made from the measured \ktwo\ amplitude by Equation~\ref{eq:smearing}. The strength of this agreement and the lack of other signals in the short-cadence photometry leads us to interpret \ktsph\,\ufreq\ and \ktoph\,\ufreq\ as the true frequencies of the signals found in the \ktwo\ data. Phase-folding the ULTRACAM and McDonald data on these two frequencies gives very similar, sawtooth-shaped lightcurves (Figure~\ref{fig:photom-phasefold-j1351}).

The amplitudes of the two signals are approximately equal at their fundamental frequencies. At their third harmonic (we use ``third harmonic'' to refer to $3 f_0$, where $f_0$ is the fundamental harmonic) the higher frequency signal of the two has a higher amplitude than the shorter period signal.





\section{Discussion}
\label{discussion}

\subsection{Classification as an AM CVn star}

The presence of strong, double-peaked helium lines in the spectrum of J1351, the absence of spectroscopic hydrogen, and the 92.0\,\ufreq\ RV modulations, are characteristic of AM\,CVn-type binaries. The double-peaked He~II emission originates from an accretion disc around the central white dwarf, and the spectroscopic RV period corresponds to the orbital period of the system. An orbital period this short is only possible in a binary in which both stars are degenerate or semi-degenerate. The orbital period of $15.65 \pm 0.12$\,min is between those of ES\,Cet (10.3\,minutes) and AM\,CVn itself (17.1\,minutes), making J1351 the second-shortest-period AM\,CVn-type binary to accrete via an accretion disc (see Figure~\ref{fig:histogram}). 

He II emission lines are also seen in HM\,Cnc and ES\,Cet, which are both at shorter orbital periods than J1351. In ES\,Cet the 4686\,\AA\ line is particularly strong with an equivalent width of $-80$\,\AA\ \citep[][cf.\ $-16.6$\,\AA\ for the same line in J1351]{Espaillat2005}. 
High-state AM\,CVns at longer periods than J1351 all show absorption lines rather than emission. It may therefore be the case that J1351 lies close to a transition point between emission-line systems and absorption-line systems. The other shorter-period system, V407\,Vul, is contaminated by a G star and difficult to study spectroscopically, though \citet{Steeghs2006} searched for evidence of emission lines and did not find any.

\subsection{The Nature of the Photometric Periods}
\label{nature-periods}

In comparison with other AM\,CVns that have been observed over a long baseline \citep[eg.][]{Skillman1999,Kupfer2015}, J1351 is unusually well-behaved photometrically. Only three photometric periods have been identified in J1351, all of which appear to be stable over the baseline of observations. These three signals are in good agreement with Equation~\ref{eq:freqs-j1351}. In this section we will use the data presented thus far to establish the physical origin of these signals.

The high frequency signals correspond to periods of 15.7306(3)\,minutes and 16.1121(4)\,minutes. These periods are comparable to the spectroscopic orbital period of $15.65 \pm 0.12$\,minutes. The likely interpretation is therefore that one signal is at the orbital period and the other the superhump period, an interpretation backed up by the agreement of these periods with Equation~\ref{eq:freqs-j1351}. The 15.7\,minute signal is clearly in closer agreement with the spectroscopic orbital period.
The 16.11\,minute signal disagrees with the spectroscopic period at the 3.9\,$\sigma$ level, and is most likely to be the superhump period. This interpretation fits with the pattern that orbital periods are generally shorter than superhump periods. We therefore consider this the most probable interpretation of the signals, suggesting that the disc is undergoing apsidal precession. However, we note that the possibility that they may be the other way around has not been conclusively ruled out. The $\chi^2$ values of sinusoidal fits to the RV data are 56.2 and 86.0 with 35 degrees of freedom.

This interpretation means that it is the orbital period which has a strong third harmonic ($3 f_0$). 
This may be related to the 3:1 resonance between the orbital period of the binary and the orbital period of a region of the disc, which is crucial to the mechanism by which the disc is driven to be eccentric.

The agreement with Equation~\ref{eq:freqs-j1351} implies that the low frequency signal of \longp\,\ufreq, corresponding to a period of $664.82 \pm 0.06$\,min, originates from  the precession period of the eccentric disc. The period is of the correct order for this interpretation; variability attributed to disc precession has been detected at a similar period (13.38\,hours) in AM\,CVn itself \citep{Patterson1993,Skillman1999}. 
The apparent stability of this signal throughout the \ktwo\ observation period is somewhat surprising given the variable nature of accretion discs. It suggests that the radius of the accretion disc remains approximately constant throughout the period of observation.

\begin{table}
	\centering
	\caption{Summary of properties of J1351, as derived from the \ktwo\ data. Periods are given in minutes.
	}
	\label{tab:properties}
	\begin{tabular}{lr}
		\hline
		Property & Value\\
		\hline
        Orbital period & $15.7306 \pm 0.0003$\\
        Superhump period & $16.1121 \pm 0.0004$\\
		Disc precession period & $664.82 \pm 0.06$\\
        \hline
        Superhump excess $\epsilon$ & $0.02425 \pm 0.00003$\\
        Estimated mass ratio $q$ & $0.111 \pm 0.005$\\
		\hline
	\end{tabular}
\end{table}

\subsection{Mass Ratio}
\label{mass-ratio-j1351}

\begin{table*} 
 \centering 
 \caption{A summary of the AM\,CVn mass ratios used for Figure~\ref{fig:mrdiag-j1351}. Where $q$ was derived by the superhump method, we recalculate it using Equation~\ref{eq:mrknigge} for the sake of consistency.} 
\label{tab:masses-j1351} 
 \begin{tabular}{lccccc} 
 \hline 
 Designation & $\epsilon$ & $q$ & $M_2$ ($M_\odot$) & Method & Reference \\ 
  &&&&& \\ 
 \hline 
SDSSJ1351-0643 & $0.024 \pm 0.000$ & $0.111 \pm 0.005$ & -- & Superhumps & 1\\ 
AM CVn$^{a}$ & $0.022 \pm 0.000$ & $0.180 \pm 0.010$ & $0.125 \pm 0.012$ & Spectroscopy & 2\\ 
HP Lib & $0.015 \pm 0.000$ & $0.074 \pm 0.007$ & -- & Superhumps & 3\\ 
CXOGBS J1751-2940 & $0.014 \pm 0.001$ & $0.070 \pm 0.007$ & -- & Superhumps & 4\\ 
CR Boo & $0.011 \pm 0.000$ & $0.058 \pm 0.008$ & -- & Superhumps & 3\\ 
KL Dra & $0.019 \pm 0.000$ & $0.092 \pm 0.006$ & -- & Superhumps & 5,6\\ 
V803 Cen & $0.011 \pm 0.003$ & $0.058 \pm 0.014$ & -- & Superhumps & 3\\ 
YZ LMi$^{b}$ & $0.009 \pm 0.000$ & $0.041 \pm 0.002$ & $0.035 \pm 0.003$ & Eclipses & 7\\ 
CP Eri & $0.009 \pm 0.001$ & $0.051 \pm 0.008$ & -- & Superhumps & 8\\ 
SDSSJ1240-0159 & -- & $0.039 \pm 0.010$ & -- & Spectroscopy & 9\\ 
SDSSJ0129+3842 & $0.009 \pm 0.005$ & $0.051 \pm 0.023$ & -- & Superhumps & 10\\ 
GP Com & -- & $0.020 \pm 0.003$ & -- & Spectroscopy & 11\\ 
SDSSJ0902+3819 & $0.005 \pm 0.002$ & $0.033 \pm 0.012$ & -- & Superhumps & 12\\ 
Gaia14aae & -- & $0.029 \pm 0.002$ & $0.025 \pm 0.001$ & Eclipses & 13\\ 
V396 Hya & -- & $0.014 \pm 0.004$ & -- & Spectroscopy & 14\\ 
\hline 
 \multicolumn{6}{p{12.5cm}}{References: [1]~Green in prep; [2]~\citet{Roelofs2006}; [3]~\citet{Roelofs2007-HST}; [4]~\citet{Wevers2016}; [5]~\citet{Wood2002}; [6]~\citet{Ramsay2010}; [7]~\citet{Copperwheat2011}; [8]~\citet{Armstrong2012}; [9]~\citet{Roelofs2005a}; [10]~\citet{Kupfer2013}; [11]~\citet{Marsh1999}; [12]~\citet{Kato2014}; [13]~\citet{Green2018}; [14]~\citet{Kupfer2016}.}\\ 
\hline 
 \multicolumn{6}{p{12.5cm}}{$^{a}$ The tabulated $q$ for AM\,CVn itself is derived from spectroscopy; its superhump excess gives $q = 0.101 \pm 0.005$ by Equation~\ref{eq:mrknigge} \newline $^{b}$ The tabulated $q$ for YZ\,LMi is derived from eclipse photometry; its superhump excess gives $q = 0.049 \pm 0.008$ by Equation~\ref{eq:mrknigge}}\\ 
\hline 
 \end{tabular} 
 \end{table*}

\begin{figure}
\includegraphics[width=\columnwidth]{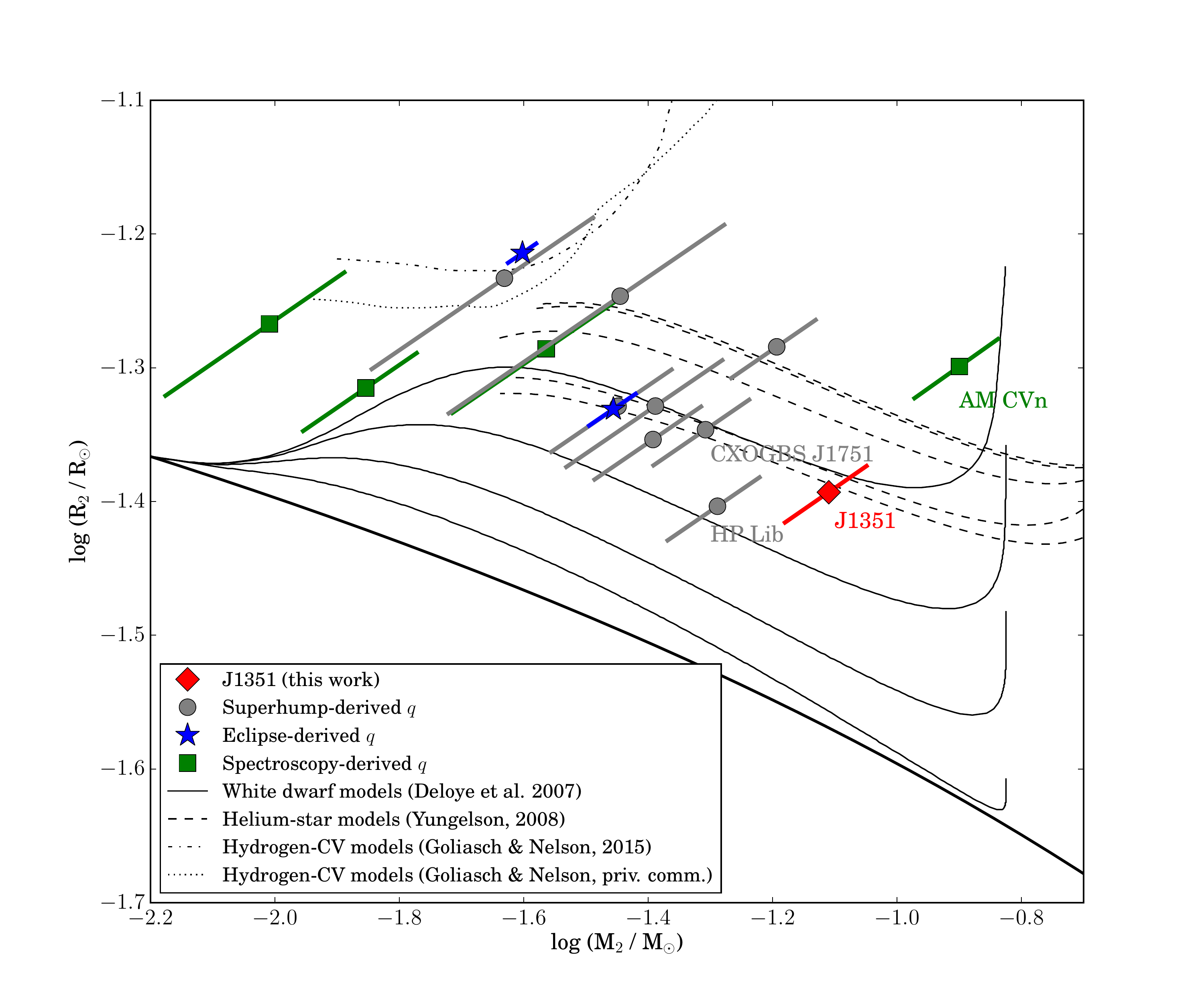}
\caption{Measured donor masses or mass ratios for a sample of AM\,CVns, compared to predicted $M$--$R$ tracks for donors in AM\,CVns descended from three proposed formation channels. Error bars are diagonal because of the strong constraint on mean density which comes from orbital period \citep{Faulkner1972a}. The thick black line shows the $M$--$R$ relation for a zero-entropy white dwarf. For systems where M$_1$ is not known, a value of $0.7 \pm 0.1$ has been assumed.
}
\label{fig:mrdiag-j1351}
\end{figure}

Taking the measurements of the orbital and superhump frequencies determined in Section~\ref{nature-periods}, we can estimate the mass ratio, $q = M_2/M_1$, of the binary from the empirical relation with superhump excess \citep{Knigge2006}
\begin{equation}
q(\epsilon) = (0.114 \pm 0.005) + (3.97 \pm 0.41) \times (\epsilon - 0.025)
\label{eq:mrknigge}
\end{equation}
where $\epsilon = (P_\text{sh} - P_\text{orb}) / P_\text{orb}$ is the superhump excess.
This relation gives similar results to that of \citet{Patterson2005}, but with the inclusion of uncertainties on the fit parameters. When applied to J1351 these uncertainties dominate due to the small uncertainty on $\epsilon$. We emphasise that this relation was derived for hydrogen-dominated CVs, and has not yet been well tested for AM\,CVns \citep[see eg.][]{Roelofs2006}. With this caveat, we find $\epsilon = 0.02425(3)$ which gives an estimate of $q = 0.111 \pm 0.005$.

In Figure~\ref{fig:mrdiag-j1351} we put this measurement in context with the other AM\,CVn systems. We show the donor mass and radius implied by this mass ratio, together with comparable donor properties for other known AM\,CVns, as listed in Table~\ref{tab:masses-j1351}. Where the $q$ was derived from superhump excess, we redid this calculation using the \citet{Knigge2006} relation in order to ensure consistency. For systems in which only a mass ratio is known (including J1351), we assume a primary mass $M_1 = 0.7 \pm 0.1$. Error bars are diagonal due to the tight constraint on the mean density of the donor which comes from the orbital period of the system \citep{Faulkner1972a}.

We also show model $M_2$--$R_2$ tracks for three evolutionary channels which may contribute to the AM\,CVn population. A brief overview of these channels is as follows. In the white dwarf donor channel \citep{Paczynski1967,Deloye2007}, the system evolves from a detached binary consisting of two white dwarfs. These white dwarfs inspiral due to gravitational wave radiation until they are close enough to begin mass transfer, at orbital periods of around 5--10\,minutes. The white dwarf which becomes the donor may not be completely degenerate. The helium star donor channel \citep{Savonije1986,Iben1987,Yungelson2008} is similar, but in this case the progenitor system consists of one white dwarf and one core helium-burning star, the atmosphere of which has been stripped. In a system descended through this channel, the additional thermal support within the donor will give it a larger radius for a given mass. In the evolved CV channel \citep{Tutukov1985, Podsiadlowski2003,Goliasch2015}, the AM\,CVn is descended from a CV with an evolved donor. As the atmosphere of the donor is stripped away and its helium core is revealed, the transferred matter becomes helium-dominated. This channel predominantly forms AM\,CVns with long orbital periods, and is thought to make only a negligible contribution to the population of AM\,CVns with orbital periods of less than 30 minutes \citep{Nelemans2004,Goliasch2015}.

From Figure~\ref{fig:mrdiag-j1351}, the population as a whole appears to include only donors with a significant amount of thermal support. Given the tracks shown here, this seems to favour the helium star donor channel. This is not conclusive, however: it may also be the case that the effect of irradiation of the donor star has been underestimated. It is also worth noting that mass ratios derived by the superhump relation may yet have an unknown bias in helium-dominated systems.

J1351 is in a region of parameter space that can be explained by either the white dwarf donor channel or the helium star donor channel. It is reasonably consistent with the other high-state systems HP\,Lib and CXOGBS\,J1751. AM\,CVn itself appears to be something of an outlier, as it falls in a region of parameter space that is difficult to explain by any of these formation channels, unless it is a pre-bounce system forming by the white dwarf donor channel. 


\subsection{Distance and Space Density}

After SDSS\,1908+3940 \citep{Fontaine2011}, J1351 is the second AM\,CVn to be discovered in the footprints of \Kepler\ and \ktwo. Given the rarity of AM\,CVns, this is worthy of note. A survey of AM\,CVns in 11663~square~degrees of SDSS~DR7 that was complete to a magnitude \filg$<19$ included only 4 AM\,CVns within that magnitude limit \citep[though note the total number of AM\,CVns discovered by the survey was larger]{Carter2013}. Based on this, \citet{Carter2013} estimated an AM\,CVn space density of $(5 \pm 3) \times 10^{-7} \text{pc}^{-3}$. Both high-state systems found by \Kepler\ and \ktwo\ are also within the \filg$<19$ limit. Scaling the population found by \citet{Carter2013} by the ratio of that area to the total \Kepler+\ktwo\ area up to and including Campaign 6 ($\approx 750$~square~degrees) we would expect 0.25 AM\,CVns with \filg$<19$ in the \Kepler+\ktwo\ footprint. \Kepler\ and \ktwo\ have therefore found more AM\,CVns than would be expected, but given the small numbers involved and the large uncertainty on the space density, the discrepancy is unlikely to be significant.

Also surprising is the coincidence that both these systems are high-state binaries. Including J1351, the six known high-state systems comprise only $\sim 12$~per~cent of the known AM\,CVn population. However, the sample of all known systems includes a selection bias toward outbursting systems due to transient surveys \citep[eg.][]{Levitan2013}. If we define a sample including only systems with magnitudes $<19$ \citep[the magnitude limit used by][who selected this object as a candidate white dwarf]{GentileFusillo15}, using quiescent magnitudes for outbursting systems, the bias towards outbursting systems is reduced. High-state systems become $\sim 35$--$40$~per~cent of the sample, and finding two high-state systems then becomes somewhat more probable. Note that high-state systems are expected to make up $\lesssim 2$~per~cent of the AM\,CVn population \citep{Roelofs2007}, but this is countered in a magnitude-limited sample by their brighter absolute magnitudes.

An estimate of the distance to J1351 can be made from the predicted absolute magnitudes for disc-dominated AM\,CVns calculated by \citet{Nelemans2004}. For an AM\,CVn with this orbital period, the predicted absolute magnitude would be $\approx 6$--$8$. By comparison with our apparent magnitude we estimate a distance of 130--330\,pc. This prediction assumes an AM\,CVn descended from a double white dwarf; if the donor of the system is instead descended from a semi-degenerate helium star, the mass transfer rate would be greater and hence the magnitude would be brighter, giving a larger distance. A reliable distance estimate should be given by \textit{Gaia} \citep{Gaia2016} in the near future.


\section{Conclusions}

We have presented the discovery of J1351, a system with a spectroscopic period of $15.65 \pm 0.12$~minutes that was discovered using \ktwo\ data. The spectrum, orbital period, and lightcurve of this object are consistent with a classification as an AM\,CVn-type binary. This makes J1351 one of only a small number of known disc-accreting high-state AM\,CVn-type systems, and the second discovered using \Kepler\ or \ktwo\ photometry.

J1351 has several visible photometric periods, including a disc precession period at $664.82 \pm 0.06$\,minutes, a signal at $15.7306 \pm 0.0003$\,minutes which is in agreement with its orbital period, and a signal at $16.1121 \pm 0.0004$\,minutes which we identify as the superhump period. Using the empirical relation of \citet{Knigge2006}, we can estimate the mass ratio of the binary as $q = M_2/M_1 = 0.111 \pm 0.005$. This mass ratio is presented with the caveat that the relationship between superhump excess and mass ratio may not be reliable for helium-dominated binaries.

As a short-period AM\,CVn, J1351 is likely to be a bright emitter of low-frequency gravitational waves. Further study may provide the mass estimates required to quantify its emission.
The presence of a photometric signature of the orbital period provides an exciting opportunity to track the period evolution of the system over the next few years, providing a constraint on the system component masses.  
However, the alignment of this period with a nightly alias of the superhump period means such efforts will likely require multi-site observations if performed from the ground. 
The system has been re-observed by \ktwo\ in Campaign 17 in short-cadence (58.8\,s) mode, allowing an opportunity to revisit this analysis and providing a longer baseline with which to constrain the period evolution.

This work highlights the fact that AM\,CVns have photometric variability on both short and long timescales. Sustained, high-speed photometry can yield a great deal of information on the nature of the system.

\section*{Acknowledgements}

The authors are grateful to Thomas Kupfer and Gavin Ramsay for their help in assembling Table~\ref{tab:amcvns-j1351}, as well as to Mukremin Kilic, whose Guest Observer program GO6003 ensured the discovery of J1351.

MJG acknowledges funding from an STFC studentship via grant ST/N504506/1. 
Support for this work was provided by NASA through Hubble Fellowship grant \#HST-HF2-51357.001-A, awarded by the Space Telescope Science Institute, which is operated by the Association of Universities for Research in Astronomy, Incorporated, under NASA contract NAS5-26555. 
TRM, DTHS and EB acknowledge STFC via grants ST/L000733/1 and ST/P000495/1. 
KJB acknowledges support from NSF grant AST-1312983.
VSD, SPL, and ULTRACAM are funded by STFC via consolidated grant ST/J001589. 
The research leading to these results has received funding from the European Research Council under the European Union's Seventh Framework Programme (FP/2007-2013) / ERC Grant Agreement n. 320964 (WDTracer).

This paper includes data collected by the \ktwo\ mission. Funding for the \ktwo\ mission is provided by the NASA Science Mission directorate.
This paper includes data taken at The McDonald Observatory of The University of Texas at Austin.
This research is based in part on observations obtained at the Southern Astrophysical Research (SOAR) telescope (NOAO Prop. ID: 2017A-0212; PI: J. J. Hermes), which is a joint project of the Minist\'{e}rio da Ci\^{e}ncia, Tecnologia, e Inova\c{c}\~{a}o (MCTI) da Rep\'{u}blica Federativa do Brasil, the U.S. National Optical Astronomy Observatory (NOAO), the University of North Carolina at Chapel Hill (UNC), and Michigan State University (MSU). 
It is also based on observations collected at the European Organisation for Astronomical Research in the Southern Hemisphere.

This publication made use of the packages \textsc{pamela}, \textsc{molly}, \textsc{numpy}, \textsc{matplotlib}, \textsc{astropy}, and \textsc{scipy}. 




\bibliographystyle{mnras}
\bibliography{refs} 





\appendix

\section{Summary of Published AM\,CVn Orbital Periods}

\begin{table*} 
 \centering 
 \caption{A summary of published AM\,CVn orbital periods. ``Method'' describes the method by which the orbital period was determined: either by timing of eclipses, spectroscopic RVs, or photometric variability.} 
\label{tab:amcvns-j1351} 
 \begin{tabular}{lcccccc} 
 \hline 
 Designation & Coordinates &  \multirow{2}{6em}{\centering Orbital\\Period (min)} & Category & Method & \multirow{2}{6em}{\centering \ktwo\ \\ Campaign} & Reference\ \\ 
  &&&&&& \\ 
 \hline 
HM Cnc & 08:06:22.84 +15:27:31.5 & 5.4 & Direct impact & Spectroscopic & -- & 1,2 \\ 
V407 Vul & 19:14:26.09 +24:56:44.6 & 9.5 & Direct impact & Spectroscopic & -- & 3,4,5 \\ 
ES Cet$^{a}$ & 02:00:52.17 -09:24:31.7 & 10.3 & High state & Photometric & -- & 6,7 \\ 
SDSSJ1351-0643 (J1351) & 13:51:54.46 -06:43:09.0 & 15.7 & High state & Spectroscopic & 6,17 & 8 \\ 
AM CVn & 12:34:54.60 +37:37:44.1 & 17.1 & High state & Spectroscopic & -- & 9 \\ 
SDSSJ1908+3940$^{b}$ & 19:08:17.07 +39:40:36.4 & 18.2 & High state & Spectroscopic & -- & 10 \\ 
HP Lib & 15:35:53.08 -14:13:12.2 & 18.4 & High state & Spectroscopic & 15 & 11,12 \\ 
PTF1J1919+4815 & 19:19:05.19 +48:15:06.2 & 22.5 & Outbursting & Eclipses & -- & 13 \\ 
CXOGBS J1751-2940 & 17:51:07.6 -29:40:37 & 22.9 & High state & Photometric & -- & 14 \\ 
CR Boo & 13:48:55.22 +07:57:35.8 & 24.5 & Outbursting & Photometric & -- & 15,16,17 \\ 
KL Dra & 19:24:38.28 +59:41:46.7 & 25 & Outbursting & Photometric & -- & 18 \\ 
V803 Cen & 13:23:44.54 -41:44:29.5 & 26.6 & Outbursting & Spectroscopic & -- & 11 \\ 
PTF1J0719+4858 & 07:19:12.13 +48:58:34.0 & 26.8 & Outbursting & Spectroscopic & -- & 19 \\ 
YZ LMi & 09:26:38.71 +36:24:02.4 & 28.3 & Outbursting & Eclipses & -- & 20 \\ 
CP Eri & 03:10:32.76 -09:45:05.3 & 28.4 & Outbursting & Photometric & -- & 21 \\ 
PTF1J0943+1029 & 09:43:29.59 +10:29:57.6 & 30.4 & Outbursting & Spectroscopic & -- & 22 \\ 
V406 Hya & 09:05:54.79 -05:36:08.6 & 33.8 & Outbursting & Spectroscopic & -- & 23 \\ 
PTF1J0435+0029 & 04:35:17.73 +00:29:40.7 & 34.3 & Outbursting & Spectroscopic & -- & 22 \\ 
SDSSJ1730+5545 & 17:30:47.59 +55:45:18.5 & 35.2 & No outbursts & Spectroscopic & -- & 24 \\ 
SDSSJ1240-0159 & 12:40:58.03 -01:59:19.2 & 37.4 & Outbursting & Spectroscopic & 10 & 25 \\ 
SDSSJ0129+3842 & 01:29:40.06 +38:42:10.5 & 37.6 & Outbursting & Spectroscopic & -- & 26 \\ 
SDSSJ1721+2733$^{c}$ & 17:21:02.48 +27:33:01.2 & 38.1 & Outbursting & Not specified & -- & 27 \\ 
SDSSJ1525+3600 & 15:25:09.58 +36:00:54.6 & 44.3 & Outbursting & Spectroscopic & -- & 26 \\ 
SDSSJ0804+1616 & 08:04:49.49 +16:16:24.8 & 44.5 & No outbursts & Spectroscopic & -- & 28 \\ 
SDSSJ1411+4812$^{d}$ & 14:11:18.31 +48:12:57.6 & 46 & No outbursts & Spectroscopic & -- & 29 \\ 
GP Com & 13:05:42.43 +18:01:04.0 & 46.5 & No outbursts & Spectroscopic & -- & 30,31 \\ 
SDSSJ0902+3819 & 09:02:21.36 +38:19:41.9 & 48.31 & Outbursting & Spectroscopic & -- & 32 \\ 
Gaia14aae & 16:11:33.97 +63:08:31.8 & 49.7 & Outbursting & Eclipses & -- & 33,34 \\ 
SDSSJ1208+3550 & 12:08:41.96 +35:50:25.2 & 52.96 & No outbursts & Spectroscopic & -- & 26 \\ 
SDSSJ1642+1934 & 16:42:28.08 +19:34:10.1 & 54.2 & No outbursts & Spectroscopic & -- & 26 \\ 
SDSSJ1552+3201 & 15:52:52.48 +32:01:50.9 & 56.3 & No outbursts & Spectroscopic & -- & 35 \\ 
SDSSJ1137+4054$^{d}$ & 11:37:32.32 +40:54:58.3 & 59.6 & No outbursts & Spectroscopic & -- & 36 \\ 
V396 Hya & 13:12:46.93 -23:21:31.3 & 65.1 & No outbursts & Spectroscopic & -- & 37 \\ 
\hline 
 \multicolumn{7}{p{18cm}}{References: [1]~\citet{Israel2002}; [2]~\citet{Roelofs2010}; [3]~\citet{Motch1996}; [4]~\citet{Cropper1998}; [5]~\citet{Steeghs2006}; [6]~\citet{Woudt2003}; [7]~\citet{Copperwheat2011a}; [8]~This work; [9]~\citet{Nelemans2001a}; [10]~\citet{Kupfer2015}; [11]~\citet{Roelofs2007a}; [12]~\citet{Roelofs2007-HST}; [13]~\citet{Levitan2014}; [14]~\citet{Wevers2016}; [15]~\citet{Provencal1991}; [16]~\citet{Provencal1994}; [17]~\citet{Patterson1997}; [18]~\citet{Wood2002}; [19]~\citet{Levitan2011}; [20]~\citet{Copperwheat2011}; [21]~\citet{Armstrong2012}; [22]~\citet{Levitan2013}; [23]~\citet{Roelofs2006}; [24]~\citet{Carter2014a}; [25]~\citet{Roelofs2005a}; [26]~\citet{Kupfer2013}; [27]~\citet{Levitan2015}; [28]~\citet{Roelofs2009}; [29]~\citet{RoelofsPhD}; [30]~\citet{Nather1981}; [31]~\citet{Marsh1999}; [32]~\citet{Rau2010}; [33]~\citet{Campbell2015}; [34]~\citet{Green2018}; [35]~\citet{Roelofs2007b}; [36]~\citet{Carter2014b}; [37]~\citet{Ruiz2001}.}\\ 
\hline 
 \multicolumn{7}{p{17cm}}{$^{a}$ The state of ES Cet is uncertain; it may be a direct impact \citep{Espaillat2005} or a high state system \newline $^{b}$ This system was in the \Kepler\ field \newline $^{c}$ \citet{Levitan2015} cite Augusteijn, priv comm, for the period of this system \newline $^{d}$ Large uncertainty on orbital period}\\ 
\hline 
 \end{tabular} 
 \end{table*}

Around 50 AM\,CVn systems are currently known. 49 were included in the most recently published count \citep[see][and the references therein]{Breedt2015}. 

Of the known systems, we are aware of 32 which have published orbital periods (we do not include systems for which only superhump periods are known). We list these systems and their orbital periods in Table~\ref{tab:amcvns-j1351}. We also specify whether the orbital period has been determined by spectroscopic RVs, by eclipse timing, or by photometric modulation. Those periods determined spectroscopically or by eclipse timing should be reliable. Periods which are only known through photometric modulation should be treated with caution, as such periods have been proved wrong by spectroscopic measurements in the past \citep[eg.][]{Roelofs2007a}.

We label those systems which were visible in past \ktwo\ Campaigns and those which will be visible in the future.

\bsp	
\label{lastpage}
\end{document}